\documentclass[twocolumn,showpacs,preprintnumbers,amsmath,amssymb,superscriptaddress]{revtex4}

\usepackage{amsmath}
\usepackage{amssymb}
\usepackage{graphicx}

\input epsf
\input rotate

\def\bea{\begin{eqnarray}}
\def\eea{\end{eqnarray}}
\def\ben{\begin{equation}}
\def\een{\end{equation}}
\def\benu{\begin{enumerate}}
\def\enu{\end{enumerate}}

\def\sss{\scriptscriptstyle\rm}

\def\1var{(\bx_1...\bx\N)}

\def\half{\frac{1}{2}}

\def\br{{\bf r}}

\def\bx{{x}}

\def\N{_{\sss N}}
\def\H{_{\sss H}}

\def\ext{_{\rm ext}}

\def\sph_int{ {\int d^3 r}}

\def\LER{_{\sss LER}}

\def\Re{{\rm Re}}
\def\Im{{\rm Im}}

\def\infintd3r{ \int_{-\infty}^\infty d^3r\,}
\def\intd3r{ \int d^3r\,}

\def\laplace1d{\frac{d^2}{dx^2}}
\def\plaplace1d{\frac{d^2}{d{x'}^2}}

\def\padr2{\frac{\partial^2}{\partial r^2}}

\def\t{\theta}
\def\E{{\cal E}}
\def\d{*}

\def\H{{\hat{H}}}

\begin{document}

\title{{Hohenberg-Kohn theorem for the lowest-energy resonance of unbound systems}}
\author{Adam Wasserman}
\affiliation{Department of Chemistry and Chemical Biology, Harvard
University, 12 Oxford St., Cambridge MA 02138, USA}
\author{Nimrod Moiseyev}
\affiliation{ITAMP -- Institute for Theoretical Atomic and
Molecular Physics, Harvard-Smithsonian Center for Astrophysics,
Cambridge, MA 01238} \affiliation{Schulich Faculty of Chemistry
and Minerva Center for Non Linear Physics, Technion - Israel
Institute of Technology, Haifa 32000, Israel}

\begin{abstract}

We show that under well-defined conditions the Hohenberg-Kohn
theorem (HKT) can be extended to the lowest-energy resonance of
unbound systems. Using the Gel'fand Levitan theorem, the extended
version of the HKT can also be applied to systems that support a
finite number of bound states. The extended version of the HKT
provides an adequate framework to carry out DFT calculations of
negative electron affinities.

\end{abstract}

\maketitle

 Most ground-state properties of electronic systems can now be
calculated from first principles via Density Functional Theory
(DFT) \cite{Kb99}. When the ground state of an $N$-electron system
is not bound, strict application of DFT with the {\em exact}
exchange-correlation functionial should yield results that are
identical to those of the $(N-M)$-electron system, where $N-M$ is
the maximum number of electrons that the external potential can
bind.
But in such cases, rather than the absolute ground state one is
often interested in resonant states (long-lived metastable
states), even if not eigenstates of the Hamiltonian. It is the
lowest-energy resonance (LER) that plays the role of the ground
state in the sense that, during its lifetime, it best represents
the physical state of the $N$-electron system. In fact, the LER is
the localized ``ground state" of the non-hermitian operator that
is obtained by complex-scaling the coordinates of the original
$N$-electron Hamiltonian by an appropriate phase factor (for
complex scaling techniques, see refs.\cite{R82,J82,M98}). The
energy $\E\LER$ and inverse lifetime $\Gamma\LER$ of the LER are
given respectively by the real and imaginary parts of its complex
eigenvalue $E\LER$. One can associate a complex density
$n_{\theta}(\br)$ to it, and intuition suggests that the
Hohenberg-Kohn theorem (HKT) that provides the foundation of DFT
\cite{HK64} can be extended along the following lines: the complex
density $n_\t(\br)$ associated with the LER uniquely determines
the $\t$-scaled external potential, and all properties of the LER
(in particular $\E\LER$ and $\Gamma\LER$) are therefore
functionals of $n_\t(\br)$. History teaches us to be watchful,
however, since the HKT has proven elusive when attempting to
depart from the ground state. It has been shown to hold for the
lowest-energy state of any given symmetry \cite{GL76, Bb79}, with
the resulting functionals depending on the particular quantum
numbers corresponding to each symmetry. More importantly, the lack
of an HKT for excited states was recently demonstrated\cite{GB04},
i.e. excited-state densities, in general, do not uniquely
determine the external potential. We are faced here with a rather
different problem, since the LER is an eigenstate of the
complex-scaled Hamiltonian rather than the unscaled one. The
simplest example is the $^3P$ resonance of H$^-$ (the
lowest-energy state for that symmetry \cite{BC94}), for which
ground-state DFT would predict its energy, had nature not made it
unbound.

 In spite of the fact that the HKT  holds
 only for the ground state\cite{DFT-NM}, attempts have been carried
  out to use time-indepedent DFT for the calculation of excited
  energy levels \cite{DFT-excited}, and even very recently
  for resonances that can
be regarded as excited states where the widths (inverse lifetimes)
are not equal to zero, but still small\cite{DFT-res}. Whereas our
aim is similar in spirit to that of ref.\cite{DFT-res}, we focus
our attention on the LER of unbound systems.





Complex-coordinate scaling is a well-developed technique to
characterize resonant states: upon multiplying all electron
coordinates of the Hamiltonian by a phase factor $e^{i\t}$, the
complex-scaled Hamiltonian $\hat{H}_\t$ has right and left
eigenvectors, denoted by  $|\Psi_\t\rangle$ and
$\langle\Psi_\t^{\d}|$ respectively, at: (a) bound states of the
original ($\theta=0$, hermitian) Hamiltonian, corresponding to the
same energy eigenvalues; (b) continuum states of the original
Hamiltonian; the respective eigenvalues are rotated into the
lower-half of the complex energy plane by an angle of $2\t$; and
(c) resonant states, with $\theta$-independent eigenvalues.
The complex-scaled resonance eigenfunctions are exponentially
localized in the interaction region, whereas the continuum
eigenfunctions almost vanish there \cite{ido-avner-NM}. We emphasize that the variational theorem for
non-hermitian quantum mechanics has been developed only for
resonances and {\it not} for the continua\cite{M82}.

The original Hohenberg-Kohn proof \cite{HK64} is based on the
minimum principle for the ground-state energy and on the fact that
the $N$-electron density operator $\hat{n}(\br)=\sum_{i=1}^N \delta(\br-\hat{\br}_i)$ couples linearly with
the external potential $v\ext(\br)$, i.e. that
one can always write a static $N$-electron Hamiltonian as: \ben
\H=\hat{T}+\hat{V}_{ee}+\int {d\br \hat{n}(\br)v\ext(\br)}~~, \een
where $\hat{T}$ is the $N$-electron kinetic-energy operator, and
$\hat{V}_{ee}$ is the electron-electron repulsion. This linear
coupling is of course maintained upon scaling (${\bf z}=\br
e^{i\t}$), \ben \H_\t=\hat{T}_\t+\hat{V}_{ee,\t}+\int {d{\bf z}
\hat{n}({\bf z}) v\ext({\bf z})}~~, \label{e:H_theta}\een where
$\hat{T}_\t=e^{-2i\t}\hat{T}$,
$\hat{V}_{ee,\t}=e^{-i\t}\hat{V}_{ee}$ for Coulomb interactions,
and $\hat{n}({\bf z})=\exp(-i3\theta)\hat{n}(\br)$\cite{delta};
but the minimum principle no longer holds: the complex variational
principle\cite{M82} guarantees stationarity at all the resonant
eigenfunctions of $\H_\t$, but not minimality at any of them. The
main result of this paper is the realization that minimality at
the LER is generally true for unbound systems, and, as a
consequence, a practical analog of the HKT can be established.


{\em {\underline {Minimality at the LER.}}} Start with a simple
case to motivate our statements. Consider a single electron moving
in the one-dimensional potential \ben v\ext(x)=\left(\half
x^2-\alpha\right)e^{-\beta x^2}~~,x>0,\label{e:potential}\een
($v\ext\to\infty$ for $x\leq 0$), and choose $\alpha$ and $\beta$
so that $v\ext$ has no bound states. Fig.1 shows this potential,
along with the complex energies and magnitude squared of five
resonance wavefunctions obtained via complex-scaling for an
appropriate choice of $\alpha$ and $\beta$. The LER energy for
such choice is
$\E\LER=\Re\langle\Psi\LER^{\d\t}|\H_\t|\Psi\LER^\t\rangle=0.62{\rm
a.u.}$  We ask whether the expectation value of $\H_\t$ for an
arbitrary trial square-integrable function $\Phi_{\sss trial}$ can
have a real part that is less than $\E\LER$. Choose for example
$\Phi_{\sss trial}(x)=Cxe^{-\gamma x^2}$ and set $C$ so that
$\Phi_{\sss trial}^\t=\Phi_{\sss trial}(xe^{i\t})$ is properly
normalized. We show in Fig.2 the energy $\E_{\sss
trial}=\Re\langle\Phi_{\sss trial}^{\d\t}|\H_\t|\Phi_{\sss
trial}^\t\rangle$ as a function of $\gamma$ and note that it is
above $\E\LER$ for all $\gamma$. According to the bounds derived
by Davidson et. al. \cite{DEM86} for resonance positions and
widths, there is no reason to expect this to be always the case,
since the most one can say about the exact complex eigenvalue at a
resonance is that it lies within a circle of radius determined by
the complex variance associated with the trial wavefunction. But
the LER of unbound systems is a special resonance since no
$\t$-independent eigenvalues of $\H_\t$ exist below it. A local
minimum at the LER must also be a global one. Nothing
 guarantees, however, that the energy at the LER is a local minimum, rather than a local maximum, or saddle point.

\begin{figure}
\begin{center}
\epsfxsize=80mm \epsfbox{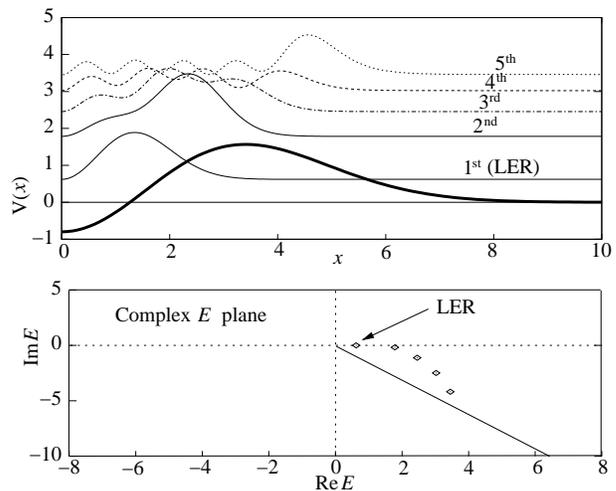}
\caption{Top panel: potential of Eq.(\ref{e:potential}) (thick line) and magnitude squared of  5 resonance
 wavefunctions ($\alpha=0.8$ and $\beta=0.1$). Bottom panel: Energy spectrum of $\H_\t$ for $\t=0.5$.
 There are no bound states. The continuum branch cut is rotated by $-2\t$ with respect to the real axis,
 and 5 resonances are clearly exposed.}
\label{f:fig1}
\end{center}
\end{figure}

\begin{figure}
\begin{center}
\epsfxsize=80mm \epsfbox{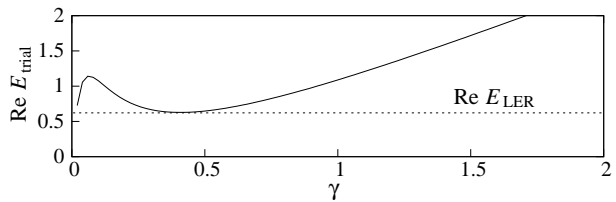}
\caption{The real part of the trial energy is above the LER energy for all values of $\gamma$ (see text).}
\label{f:fig1}
\end{center}
\end{figure}

We now argue that the result observed in Fig.2 for our test
example is in fact {\em usually} the case, also for $N$ electrons, and discuss the
plausibility of this statement for a special subset of trial
functions: those that are localized in the region where the
resonance wavefunctions, and in particular the LER, have a high amplitude (see our
comment above on the localization of the resonances in the
interaction region, whereas the rotating continuum states are not localized).
 To be specific, restrict the discussion to trial functions that satisfy:
 \ben
 |\langle\Psi\LER^{\d\t}|\Phi_{\sss trial}\rangle|^2 > \half~~.
 \label{e:half_condition}
 \een
 The same condition was employed before in ref.\cite{DEM86} to derive upper and lower bounds for resonances.
 In such cases, $\Phi_{\sss trial}$ may
 be expanded in terms of only resonance states.
 There are no bound states to populate, and the overlap with $\t$-scaled
 continuum eigenstates is negligible.
 (On the fact that the resonances serve as an almost complete set, see
 Ref.\cite{Ryaboy-NM}).
 Then, to a good approximation,
\ben E_{\sss trial}=E\LER\left[1+\sum_{k\neq{\sss LER}}
\langle\Psi_k^{\d\t}|\Phi_{\sss trial}^\t\rangle^2\left(E_k/
E\LER-1\right)\right ]~~,\label{com-exp} \een where the
$\Psi_k^\t$ are resonance eigenfunctions of $\H_\t$ with complex
eigenvalues $E_k=\E_k-\frac{i}{2}\Gamma_k$. We conclude that if
\begin{eqnarray}
\nonumber
&&\sum_{k\neq{\sss LER}}\left[\Re\langle\Psi_k^{\d\t}|\Phi_{\sss trial}^\t\rangle^2(\E_k-\E\LER)+\right.
\\&&~~~~~~~~~~~\left.+\half\Im\langle\Psi_k^{\d\t}|\Phi_{\sss trial}^\t\rangle^2(\Gamma_k-\Gamma\LER)\right]>0~~,
\label{e:main_ineq}\end{eqnarray}
then
\ben\E_{\sss trial}>\E\LER\label{e:minimum_principle}\een
The inequality of Eq.(\ref{e:main_ineq}) would be obviously true if we were dealing with bound states, since
 then all the inverse lifetimes would be zero, all the real parts of the overlaps would be positive,
 and $\E_k-\E\LER>0$ for all $k$, by definition of the LER.

Eq.(\ref{e:main_ineq}), and therefore Eq.(\ref{e:minimum_principle}),
also hold true for resonances because:
(1) Normalization of the trial function, $\sum_{k}
\Re\langle\Psi_k^{\d\t}|\Phi_{\sss trial}^\t\rangle^2 = 1$,
together with the condition given by Eq.(\ref{e:half_condition})
and the fact that $\E_k-\E\LER$ is positive for all $k>1$ ($k=1$
denotes the LER), lead to  $\sum_{k\neq 1}\Re\langle
\Psi_k^{\d\t}|\Phi_{\sss trial}^\t\rangle^2(\E_k-\E\LER) > 0$; and
(2) Normalization also requires
$\sum_{k}\Im\langle \Psi_k^{\d\t}|\Phi_{\sss trial}^\t\rangle^2 =
0$, so if $\Gamma_{k\ne 1}-\Gamma\LER$ is a smooth function of
$k$, as is usually the case,
 then the second term of Eq.(\ref{e:main_ineq})
is expected to be smaller than the first one.

 We illustrate all this in Fig.3 for our one-electron toy example. The
two conditions discussed above are shown to hold in the region
where an expansion in terms of resonance eigenstates is adequate
(shaded region in the figure).  Based on the result of Fig.2
showing that $\E_{\sss trial}>\E\LER$ even outside this range,
  we infer that the LER energy is embedded
  inside a left half-circle in the complex energy plane: the left half of the circle where the exact solution is embedded according to
  Ref.\cite{DEM86}.
We summarize it by saying that under the conditions stated above,
the energy of the LER, $\Re E\LER={\cal E}\LER$, which is
associated with the real part of the complex eigenvalue of the
non-hermitian hamiltonian, satisfies the following modified
complex variational principle:
\begin{equation}
{\cal E}\LER = \mathop{\rm min}_{\Psi_\t}
\Re\langle\Psi_\t^{\d}|\H_\t|\Psi_\t\rangle \label{e:min_LER}
\end{equation}

{\em{\underline {Hohenberg-Kohn theorem.}}}
\begin{figure}
\begin{center}
\epsfxsize=80mm \epsfbox{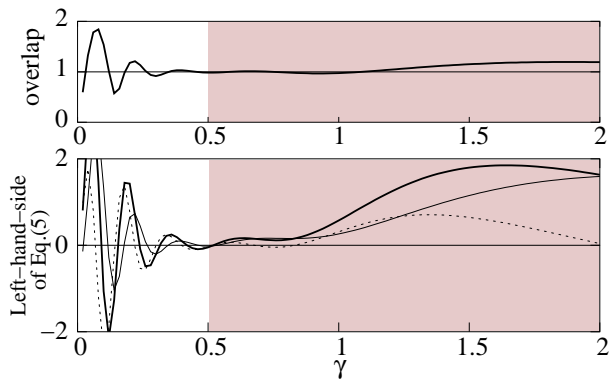}
\caption{Top pannel: Total overlap $S$ of a trial function
$\Phi_\gamma^\t(x)=Cxe^{i\t}e^{-\gamma x^2e^{2i\t}}$ with the resonance
eigenfuctions $\Psi_k^\t$ of $\H_\t$: $S=\left|\sum_{k=1}^5\langle\Psi_k^{\d\t}|\Phi_\gamma^\t\rangle^2\right|$.
 The expansion is adequate in the shaded region. Deviations from $S=1$ for large $\gamma$  diminish when broader resonances are included in the expansion
 \cite{Ryaboy-NM}.
  Bottom panel: left-hand-side of Eq.(\ref{e:main_ineq})
 (thick line), along with the contribution to it from the real part of the overlaps (solid) and the imaginary part
 (dotted).}
\label{f:fig1}
\end{center}
\end{figure}
Having established the plausibility of
Eq.(\ref{e:minimum_principle}) for trial functions that can be
expanded in terms of resonance wavefunctions, an analog of the Hohenberg-Kohn
theorem follows.
 Two potentials $v\ext^{[1]}$ and $v\ext^{[2]}$ that do not support any bound state, and differ by more than a constant,
  {\em cannot} yield the same LER-density $n_\t(\br)=\langle\Psi\LER^{\d\t}|\hat{n}(\br e^{i\t})|\Psi\LER^\t\rangle$.
  To see this, assume that the two potentials could in fact give rise to the same LER-density:
\ben
\langle\Psi\LER^{\d[1]}|\hat{n}(\br e^{i\t})|\Psi\LER^{[1]}\rangle=\langle\Psi\LER^{\d[2]}|\hat{n}(\br e^{i\t})|\Psi\LER^{[2]}
\rangle~~\label{e:equal_dens},
\een
where $\Psi\LER^{[j]}$ is the LER-eigenstate of the Hamiltonian of Eq.(\ref{e:H_theta}) with external potential
$v\ext^{[j]}$. Using $\Psi\LER^{[1]}$ as a trial function to estimate the lowest-energy eigenvalue of $\H_\t^{[2]}$,
 we get by virtue of Eqs.(\ref{e:minimum_principle}) and (\ref{e:equal_dens})
  that $\Re\langle\Psi\LER^{\d[1]}|\hat{T}_\t+\hat{V}_{ee,\t}|\Psi\LER^{[1]}\rangle>
  \Re\langle\Psi\LER^{\d[2]}|\hat{T}_\t+\hat{V}_{ee,\t}|\Psi\LER^{[2]}\rangle$.
  But the opposite result is obtained by employing $\Psi\LER^{[2]}$ as a trial function
  to estimate the lowest-energy eigenvalue of $\H_\t^{[1]}$. We conclude that the original assumption
   of Eq.(\ref{e:equal_dens}) is impossible if $v\ext^{[1]}$ and $v\ext^{[2]}$ differ by more than a constant.


 To see the problem from a
different perspective,
   we now examine the Levy-Lieb \cite{L79,L83} constrained search algorithm
in the present context. The LER state is the one that, among all
the normalized wavefunctions that make the complex energy
$\langle\Psi_\t^\d|\H_\t|\Psi_\t\rangle$ stationary, minimizes the
expectation value of $\Re\langle\Psi_\t^\d|\H_\t|\Psi_\t\rangle$.
  Following Levy, we perform the minimization in two steps,
first constraining the search among all the wavefunctions yielding
a prescribed complex density, $\{\Psi_\t \mapsto n_\t\}$, and then
among all possible complex densities, $\{n_\t\}$. The energy of
the LER is then given by: \ben \E\LER=\mathop{\rm min}_{n_\t}
\Re\left[\int d{\bf z}~n_\t({\bf z}) v\ext({\bf
z})+{F_\t[n_\t]}\right]~~s.t.\left\{{\rm
c1}\right\},\label{e:N0}\een
\ben \Re{F_\t[n_\t]}= \mathop{\rm min}_{\Psi_\t \mapsto n_\t}
\left[\Re
\langle\Psi_\t^\d|\hat{T}_\t+\hat{V}_{ee,\t}|\Psi_\t\rangle\right]~~s.t.\left\{{\rm
c2}\right\}\label{e:N1}\een
 and
constraints ${\rm c1}$ and ${\rm c2}$ are as discussed before:
\begin{eqnarray} &{\rm c1:}& \int d{\bf z} n_\t({\bf z})=N \label{e:N2}\\
&{\rm c2:}&
\frac{\delta}{\delta\Psi_\t}{\langle\Psi^\d_\t|\H_\t|\Psi_\t\rangle}=0\label{e:N3}
\end{eqnarray}

In spite of the formal resemblance of Eq.(\ref{e:N0}) with
the density-variational principle that serves as a starting point
to derive Kohn-Sham equations, condition c2 makes of this a very
different problem. It introduces a seemingly very complicated
explicit dependence of $F_\t[n_\t]$ on $v\ext$, preventing proof
of the HKT analog.

But we now invoke Eq.(\ref{e:min_LER}). According to it, constraint
 c2 can be lifted altogether. The resulting (unconstrained) search of Eq.(\ref{e:N1})
defines a universal functional $\Re F_\t[n_\t]$, just as in the
ground-state case. There is no explicit dependence of $\Re
F_\t[n_\t]$ on $v\ext$, and the HKT analog is established.

To access the lifetime of the LER, denote by $\bar{\Psi}_\t$ the
wavefunction that, within the set of functions yielding $n_\t$,
minimizes the real part of
${\langle\Psi_\t^\d|\H_\t|\Psi_\t\rangle}$ subject to the
normalization constraint c1. It is a functional of $n_\t$,
$\bar{\Psi}_\t[n_\t]$.
If we further define $\Im F_\t[n_\t]$ as
the imaginary part of: \ben
{F_\t[n_\t]}={\langle\bar{\Psi}^\d_\t[n_\t]|\hat{T}_\t+\hat{V}_{ee,\t}
|\bar{\Psi}_\t[n_\t]\rangle}~~, \label{e:N6}\een then the inverse
lifetime of the LER is given by the imaginary part of the sum of
$F_\t[n_\t]$ and $\int d{\bf z} n_\t({\bf z})v\ext({\bf z})$.

When apart from being the resonance of lowest energy, the LER is
also the resonance of longest lifetime, a typical case (e.g. our
toy example), then Eq.(\ref{e:N0}) can be subsumed by a
two-component minimization yielding at the same time the energy
$\E$ and inverse lifetime $\Gamma=\hbar/\tau$ of the LER:
\begin{eqnarray} \left(\begin{array}{c}
\E\\
\Gamma
\end{array}\right)_{\sss LER} =\mathop{\rm
min}_{n_\t}\left(\begin{array}{c}
\Re\\-2\Im\end{array}\right)\left[\int{d{\bf z}~n_\t({\bf
z})v\ext({\bf z})}+F_\t[n_\t]\right]
\label{e:LL_analog}\\\nonumber s.t.\{{\rm c1}\} \end{eqnarray}

We have admittedly not addressed here the two fundamental
questions that immediately arise:
(1) What is the best way to cast the complex analog of the
Kohn-Sham scheme for practical calculations? and (2) What is the
functional form of $F_\t[n_\t]$? For one electron, it is simple to
show that $F_\t[n_\t]=e^{-2i\t}F[n_\t]$, where $F[n_\t]$ is the
ground-state functional evaluated on the complex density.

Our derivation applies to the LER of unbound systems such as
negatively charged atoms or molecules. However, using the
Gel'fand-Levitan\cite{GELFAND} equation it is quite
straightforward to extend our formulation to systems that support
also bound states. It has been shown already that using the
Gel'fand Levitan equation one can remove bound states from the
spectrum and obtain an effective potential which supports
resonances only\cite{NM-Osvaldo}. However, from a numerical point
of view it might be a heavy task problem since the computation of
new effective potentials that support the same resonances as the
original problem, but not any of the $N$ bound states, requires
the often prohibitive calculation of those bound-state
wavefunctions. Our  extension of the HKT for the LER of unbound
systems holds also for atoms and molecules in the presence of
external DC or AC electric fields, since the field-free ground
(bound) state becomes a resonance state as the DC or AC fields are
turned on (for the calculation of such resonances via complex
scaling see refs. \cite{R82} and \cite{M98}).

{ \em{\underline {Negative electron affinities.}}}
We comment briefly on the computation of
negative electron affinities as measured experimentally for many
molecules via electron transmission spectroscopy \cite{JB87}. The
standard definition
 of the electron affinity is:
$A=E^{[N-1]}-E^{[N]}$ where $E^{[N-1]}$ is the ground-state energy
of the neutral molecule, and $E^{[N]}$ is the ground-state energy
of the negative ion. The latter is precisely equal to $E^{[N-1]}$
when the ion is not bound, so $A$ is zero in such cases. Confusion
arises in practice when a finite basis set used in DFT
calculations artificially binds the ion and predicts finite
(negative) values for $A$.  But the experiments measure a
different quantity: $ \tilde{A}=E^{[N-1]}-\Re E\LER^{[N]}$ and
this is not directly accessible via standard DFT calculations in
the limit of an infnite basis-set. It is nonetheless interesting
that $\tilde{A}-A$ is accurately given in many instances by the
error associated with the use of a finite basis set (see
discussion in ref.\cite{TP05}).


\textbf{Acknowledgements} Useful discussions with Eric Heller
are gratefully acknowledged.



\begin{thebibliography}{0}
\bibitem{Kb99}
W. Kohn,
Rev. Mod. Phys. {\bf 71}, 1253 (1999).
\bibitem{R82}
W. P. Reinhardt, Annu. Rev. Phys. Chem. {\bf 33}, 223 (1982).
\bibitem{J82}
B.R. Junker,
Adv. At. Mol. Phys. {\bf 18}, 207 (1982).
\bibitem{M98}
N. Moiseyev,
Phys. Rep. {\bf 302}, 211 (1998).
\bibitem{HK64}
P. Hohenberg and W. Kohn,
Phys. Rev. {\bf 136}, B 864 (1964).
\bibitem{GL76}
O. Gunnarsson and B.I. Lundqvist,
Phys. Rev. B {\bf 13}, 4274 (1976)
\bibitem{Bb79}
U. von Barth,
Phys. Rev. A {\bf 20}, 1693 (1979).
\bibitem{GB04}
R. Gaudoin and K. Burke,
Phys. Rev. Lett. {\bf 93}, 173001 (2004).
\bibitem{BC94}
S.J. Buckman and C.W. Clark,
Rev. Mod. Phys. {\bf 66}, 539 (1994).
\bibitem{DFT-NM} N.
Moiseyev, Chem. Phys. Lett. {\bf 321}, 469 (2000).
\bibitem{DFT-excited} S.M. Valone, J.F. Capitani, Phys. Rev. A {\bf 23} 2127 (1981).
\bibitem{DFT-res} M. Ernzerhof, J. Chem. Phys. {\bf 125}, 124104
(2006).
\bibitem{ido-avner-NM}
 I. Gilary, A. Fleischer and
N. Moiseyev,  Phys. Rev. A {\bf 72}, 012117 (2005).
\bibitem{delta}N. Moiseyev and L. S.Cederbaum, Phys. Rev A 72, 033605
(2005). Note that for spherically symmetric external potentials,
$d{\bf r}=4\pi r^2 dr$ and therefore the complex-scaled external
potential operator in Eq.(2) is given by $\int {d{\bf z}
\hat{n}({\bf z}) v\ext({\bf z})}=  \int {d{\bf r} \hat{n}({\bf r})
v\ext(\exp(i\theta){\bf r})}$.
\bibitem{M82}
N. Moiseyev, P.R. Certain, and F. Weinhold
Mol. Phys. {\bf 47}, 585 (1982).
\bibitem{L79}
M. Levy,
Proc. Nat. Acad. Sci. USA {\bf 76}, 6062 (1979).
\bibitem{L83}
E.H. Lieb, Int. J. Quantum Chem. {\bf 24}, 243 (1983).
\bibitem{DEM86}
E.R. Davidson, E. Engdahl, and N. Moiseyev,
Phys. Rev. A {\bf 33}, 2436 (1986).
\bibitem{Ryaboy-NM}
V. Ryaboy and N. Moiseyev,
 J. Chem. Phys., {\bf 98},  (1993).
\bibitem{GELFAND}I. M. Gel'fand and B. M. Levitan, Am Math. Soc.
Trans. {bf 1}, 253 (1951).
\bibitem{NM-Osvaldo} N. Moiseyev and O. Goschinski, Chem. Phys.
Lett., {\bf 120}, 520 (1985).
\bibitem{JB87}
K.D.Jordan and P.D.Burrow,
Chem. Rev. {\bf 87}, 557 (1987).
\bibitem{TP05}
D.J. Tozer and F. De Proft, J. Phys. Chem. A {\bf 109}, 8923
(2005).
\end{thebibliography}
\end{document}